\documentclass[aps,prb,twocolumn,twoside,superscriptaddress]{revtex4}
\usepackage{epsf}
\usepackage[intlimits]{amsmath}
\usepackage{amssymb}
\usepackage{graphicx}
\usepackage{psfrag}
\usepackage{float}

\newcommand{\ybco}{YBa$_2$Cu$_3$O$_{6+\delta}$}
\newcommand{\lco}{La$_2$CuO$_4$}
\newcommand{\lsco}{La$_{2-x}$Sr$_x$CuO$_4$}

\newcommand{\lczo}{La$_2$Cu$_{1-z}$Zn$_z$O$_4$}
\newcommand{\lsczo}{La$_{2-x}$Sr$_x$Cu$_{1-z}$Zn$_z$O$_4$}
\newcommand{\lclo}{La$_2$Cu$_{1-y}$Li$_y$O$_4$}

\newcommand{\Tc}{$T_\text{c}$}
\newcommand{\TN}{$T_\text{N}$}

\begin{document}

\title{
Magnetic order in lightly doped cuprates:\\
Coherent vs. incoherent hole quasiparticles and non-magnetic impurities
}

\author{Marijana Kir\'can}
\affiliation{Max-Planck-Institut f\"ur Festk\"orperforschung,
Heisenbergstr. 1, 70569 Stuttgart, Germany}
\affiliation{\mbox{Institut f\"ur Theorie der Kondensierten Materie,
Universit\"at Karlsruhe, Postfach 6980, 76128 Karlsruhe, Germany}}
\author{Matthias Vojta}
\affiliation{\mbox{Institut f\"ur Theorie der Kondensierten Materie,
Universit\"at Karlsruhe, Postfach 6980, 76128 Karlsruhe, Germany}}
\date{Nov 23, 2005}

\begin{abstract}
We investigate magnetic properties of lightly doped antiferromagnetic Mott insulators
in the presence of non-magnetic impurities.
Within the framework of the $t$-$J$ model 
we calculate the doping dependence of the antiferromagnetic order parameter
using self-consistent diagrammatic techniques.
We show that in the presence of non-magnetic impurities the antiferromagnetic
order is more robust against hole doping in comparison with the impurity-free host,
implying that magnetic order can re-appear upon Zn doping into lightly hole-doped
cuprates.
We argue that this is primarily due to the loss of coherence and reduced mobility
of the hole quasiparticles caused by impurity scattering.
These results are consistent with experimental data on Zn-doped \lsco.
\end{abstract}

\maketitle


\section{Introduction}

The undoped parent compounds of the high-\Tc\ superconducting cuprates,
like \lco, are antiferromagnetic (AF) insulators.
Upon introducing a small amount of holes, e.g., by substituting La with Sr in \lco,
the commensurate long-range order is rapidly suppressed and completely destroyed
at few percent of hole doping ($x_{\text{c}}\approx 2\,\%$ in \lsco).\cite{niederm}
For $2\,\%<x<5.5\,\%$, \lsco\ enters a spin-glass phase with incommensurate
spin correlations, and finally becomes superconducting for larger $x$, with
only short-range AF correlations in the superconducting state.
It is believed that a proper understanding of the magnetism is
crucial for a theoretical description of high-\Tc\ cuprates.

On the theoretical side, the suppression of AF order upon hole doping is
reasonably well understood.
\cite{horsch,belkaMag,richardTNeel,orbach,frust1,frust2}
Numerous approaches employ effective single-band models with strong correlations
to describe the copper oxide planes.
The magnetic order, present in the half-filled N\'eel state, is scrambled by the
motion of holes -- this leads to heavy hole quasiparticles (so-called magnetic or spin polarons)
and to a reduction of the magnetic order parameter, accompanied by a softening
of the spin-wave spectrum. \cite{horsch,belkaMag,richardTNeel}
(A different approach describing the destruction of AF order,
based on the idea that holes introduce strongly ferromagnetic frustrating bonds,
has been proposed in Refs.~\onlinecite{frust1,frust2}.)

Localized impurities provide an interesting tool to investigate bulk properties of
strongly correlated systems in general, and high-\Tc\ cuprates in particular.
Doping of non-magnetic impurities (i.e.~vacancies) into the parent
antiferromagnet,\cite{netonlsm,cherny,sandvik}
e.g. in \lczo,\cite{vajkZn}
leads to a gradual suppression of magnetism, with long-range order
surviving up to the percolation threshold, $z_{\text{p}}\approx 40.5\,\%$.
One expects that combining hole doping and non-magnetic
impurities would lead to a destruction of the AF order even faster
than with hole doping alone.
Surprisingly, magnetization measurements on \lsczo\
by H\"ucker {\it et al.} \cite{hueckerZn} have shown that this is not the case.
They found that introducing non-magnetic Zn impurities in lightly
doped \lsco\ ($x<4\,\%$) causes an initial {\em increase} of the N\'eel temperature, \TN,
while only an addition of Zn beyond $z \approx 10\,\%$ leads to a decrease of \TN.
Concomitantly, the decrease of \TN\ with increasing hole doping $x$ is
much weaker in Zn-doped compounds.
Recently, it has been found that such an impurity-induced re-appearance
of the AF ordering is even more pronounced
in samples with magnetic Ni impurities. \cite{machiZnNi}

The purpose of the present paper is a theoretical description of the
combined effect of mobile holes and static vacancies in lightly doped
Mott insulators.
We argue below that vacancies primarily prevent the holes from destroying the
long-range ordered magnetism; other (purely magnetic) effects are subdominant.
Starting from a magnetically ordered state, we consider the influence
of mobile holes on the host magnetism in a situation where hole quasiparticles
scatter off the static impurities.
Technically, we employ a $t$-$J$ model and the self-consistent Born approximation (SCBA)
to account for the interaction
between holes and spin waves; \cite{horsch,belkaMag}
the effect of the impurities is treated via a
self-consistent $T$-matrix approach.
We show that the destructive effect of holes on the background magnetism
is reduced
(i) when holes are less mobile or less dispersive and
(ii) when holes are less coherent,
both changes caused by the addition of non-magnetic impurities.
This results in commensurate antiferromagnetism being more robust against hole doping
for impurity-doped host systems compared to clean ones.
We obtain detailed magnetization curves as function of hole and impurity doping,
being in good agreement with the available data on \lsczo.
(Incommensurate magnetism, as occurs for hole doping $x>4\,\%$, with
much smaller ordering temperature, is not the subject of this paper.)

Before describing our theoretical approach, we briefly summarize
impurity effects in insulating quantum magnets and in cuprate materials.

\subsection{Impurities in quantum magnets}

In a quantum spin system (i.e.~a Mott insulator), which can be tuned
between an antiferromagnetic and a paramagnetic ground state,
the effect of static non-magnetic impurities (i.e.~vacancies)
is markedly different in both phases, the origin being the random Berry phases
introduced by the doping.
As mentioned above, introducing vacancies into the long-range ordered
magnet weakens the magnetic order.
At small impurity concentrations, this effect can be well captured
using linear spin-wave theory.\cite{cherny}
For a square lattice geometry, long-range magnetism persists up to
the percolation threshold\cite{bilnote} $z_{\rm p}$ -- this has been nicely
verified for \lczo\ using neutron scattering.\cite{vajkZn}
The quantum phase transition at $z_{\text{p}}$ displays
an interesting interplay of classical percolation physics and
quantum effects.\cite{sandvik,vojsch}

In contrast, removing spins in the quantum paramagnet
(which is a valence bond crystal) has been shown to create
effective magnetic moments in the vicinity of the
dopants.\cite{indmom1,indmom1b,indmom1c,indmom2,indmom3,ssmv}
Microscopically, the formation of the moments can be understood
as breaking of host valence bonds.
The impurity-induced moments are magnetically coupled via the (gapped) host
spin excitations and -- in the absence of geometric frustration -- eventually order
at very low temperatures.\cite{indmom2,indmom3}
Thus, at zero temperature the paramagnetic phase of the clean host
is replaced by a phase with weak magnetic order through doping with
non-magnetic impurities.

Besides vacancies, also magnetic impurities may be introduced into
quantum magnets, either by replacing host ions by ions of different spin (e.g.~Cu by Ni),
or inserting additional (e.g.~out-of-plane) magnetic ions.
At small doping, the physics is similar to the one in the case of vacancies,
i.e., it is dominated by the low-energy behavior of the impurity moments
(random Berry phases), but percolation physics at larger doping is absent.

\subsection{Impurities in cuprates}
\label{sec:idea}

Impurity effects in cuprate superconductors are diverse, most prominent
being the rapid suppression of superconductivity by in-plane impurities
like Zn or Ni substituting for Cu.
For hole concentrations in the superconducting regime,
it has also been shown that nominally
non-magnetic impurities like Zn or Li induce magnetic moments,
which have a staggered spatial structure around the actual impurity
site.
NMR \cite{bobroff1} shows that the impurity moments form at temperatures above 400 K, and
the formation mechanism is believed to be similar to that in an insulating paramagnet,
i.e., breaking of singlet bonds described above.
However, impurity-induced magnetic order has not been detected, the reason likely
being strong quantum effects (e.g.~Kondo screening of the moments) due to the presence
of mobile charge carriers.
(One exception is magnetic order in Co-doped \ybco \cite{keimerco} --
due to the large spin of Co quantum effects are weaker.)
Further effects of Zn and similar impurities in cuprates with larger hole doping include
magnetic in-gap states,\cite{ingap,waki}
the broadening of the gapped bulk spin excitations,
possible pinning of charge-density modulations (stripes),
the local modulation of the superconducting order parameter,
and large peaks in the density of states close to the Fermi level
(as seen e.g. in STM).

For smaller hole doping (i.e.~in the insulating re\-gi\-me),
the impurity effects on magnetism and transport have also
been studied extensively.\cite{hueckerZn,hueckerZn2,machiZnNi,ZnYBCO}
In \lsczo\ with $x\approx 3\,\%$, co-doping with Zn leads to a re-appearance of
ordered magnetism.
The relatively high ordering temperatures suggest that this is {\em not}
just the magnetism of impurity moments as described above for insulating
magnets (which would be weak, as in Mg-doped TlCuCl$_3$ \cite{mgtlcucl}).
The continuous evolution of the ordering temperature with $x$ and $z$ also
hints that the magnetism in the Zn and hole-doped material is smoothly
connected to the commensurate antiferromagnetism of the parent compound,
i.e., it is neither stripe-like nor strongly glassy.
(Note that the freezing temperature in the spin-glass state of \lsco\
is much below the ordering temperature observed in Refs.~\onlinecite{hueckerZn,hueckerZn2}.
The suppression of glassy behavior upon Zn doping is also consistent
with very recent $\mu$SR and susceptibility data.\cite{huecker05})
A natural conclusion is that the Zn impurities primarily ``undo''
the effect caused by the mobile holes on the commensurate antiferromagnetism.
Resistivity measurements in \lsczo\ indicate a decreasing hole mobility with
increasing Zn doping.
Taken together, we conclude that Zn impurities tend to hinder the coherent
motion of holes and thus prevent the magnetic order from being scrambled --
this will be the central idea of our approach below.
Other effects, namely the generation of effective staggered moments around Zn
and vacancy dilution of the magnetism (leading to percolation physics
for $z$ close to $z_\text{p}$), are clearly present, but are assumed to be
subleading (at least for small impurity concentration).

We note that an interesting dopant in \lco\ is Li (substituting for Cu) --
this provides both an extra hole and a vacancy in the CuO plane.
Studies\cite{lilco,liglass} of \lclo\ have shown that commensurate magnetism survives up to
$y=3\,\%$. Transport data indicate that the holes introduced by Li doping remain
localized also for larger $y$ (in contrast to Sr doping).
Recent experimental studies show that charge and spin dynamics in \lclo\ appear to be glassy
over a large range of doping.\cite{liglass}
Thus, the behavior of \lclo\ is different from the one of \lsczo\ with $x\!=\!z\!=\!y$,
the reason clearly being on the quantum chemistry side (e.g.~the pinning
potentials for holes being different for Li and Zn sites).
Ref.~\onlinecite{sushkovLi} proposed an antiferromagnetic cluster state
arising from Coulomb trapping of holes to explain some of the properties of \lclo,
but more theoretical studies are clearly required.

\subsection{Outline}

The remainder of the paper is organized as follows.
In Sec.~\ref{sec:effHam} we will introduce the effective model describing the interaction
between holes and spin waves.
In Sec.~\ref{sec:green} the hole and spin-wave Green's functions will be derived in
the framework of the SCBA for the impurity-free system.
Effects of the Zn impurities are discussed in Sec.~\ref{sec:impurity},
and Sec.~\ref{sec:magprop} gives the expressions for magnetic properties within
our approach.
Finally, numerical results together with a comparison to experimental data
are presented in Sec.~\ref{sec:results}.
A discussion concludes the paper.

We note that Korenblit {\it et al}. \cite{korenblit} have put forward a somewhat
different explanation for the experiments of H\"ucker {\em et al.},\cite{hueckerZn}
based on the frustration model.\cite{frust1,frust2}
They argued that Zn impurities remove some of the frustrating bonds generated
by mobile holes.
However, their calculations do not take into account the issue of reduced hole
mobility due to impurity doping; we will further comment on their results towards
the end of the paper.
Impurity effects in cuprates have also been studied within the spin-fermion
model.\cite{sfmodel} The authors do find a recovery of commensurate magnetism
in a hole-doped situation upon adding impurities; however, the simplifications
within the spin-fermion model (e.g. the absence of quantum spin waves) preclude
a comparison with experiments.


\section{The Effective Hamiltonian}
\label{sec:effHam}

In the following two sections we establish notation and -- to keep the paper self-contained --
summarize the magnetic polaron model together with its SCBA treatment in the
impurity-free situation, following Refs.~\onlinecite{horsch,belkaMag,diffeq}.

We start from the standard $t$-$J$ model on a square lattice, $H=H_t+H_J$,
and employ a representation using slave-fermion operators $f_i$
for spinless holes and Holstein-Primakoff bosons $b_i$ for spin flips away from
the N\'eel-ordered reference.
This gives\cite{kane,horsch,fulde}
\begin{eqnarray}
H_t^{\phantom{\dagger}}&=&-t\sum_{\langle i,j\rangle}
             f_i^{\phantom{\dagger}} f_j^\dagger
             (b_j^{\phantom{\dagger}}+b_i^\dagger)
      -t'\sum_{\langle\langle i,j\rangle\rangle}
             f_i^{\phantom{\dagger}} f_j^\dagger + \text{h.c.}
             , \label{kint} \\
H_J^{\phantom{\dagger}}&=&\frac{J}{2}\sum_{\langle i,j\rangle}
             \bigl(b_i^\dagger b_i^{\phantom{\dagger}} +
                   b_j^\dagger b_j^{\phantom{\dagger}} +
                   b_i^{\phantom{\dagger}} b_j^{\phantom{\dagger}} +
              b_i^\dagger b_j^\dagger\bigr) \label{heit}
\end{eqnarray}
in standard notation.
The first kinetic term in \eqref{kint} describes processes in which holes hop from one
to the neighboring site creating or annihilating spin waves;
the next-neighbor hopping term $t'$ is allows for direct hole motion
within one sublattice
(an additional term of the form $f_i f_j^\dagger b_j b_i^\dagger$ is neglected).
Values of $t/J = 3 \ldots 5$, $t'/t = -0.1 \ldots -0.2$ are relevant for cuprate materials.
In general, the Heisenberg term \eqref{heit} contains an additional
factor $f_i f_i^\dagger f_j f_j^\dagger$ which accounts for the loss of
magnetic energy due to the hole doping. On the mean-field level we have
$f_i f_i^\dagger=1-f_i^\dagger f_i=1-\delta$, where $\delta$ is
the hole concentration. Consequently, the AF exchange coupling is
renormalized according to $J\to(1-\delta)^2 J$, but for small hole doping
we can neglect this effect.

Fourier transforms of the $b$ ($f$) are defined in the reduced (magnetic)
Brillouin zone, leading to $b_{\bf q}^{\rm A}$, $b_{\bf q}^{\rm B}$
($f_{\bf k}^{\rm A}$, $f_{\bf k}^{\rm B}$)
for the two sublattices A and B.
The spin-wave part is diagonalized using a Bogoliubov transformation,
introducing new bosonic operators for spin waves,
$\alpha_{\bf q}$ and $\beta_{\bf q}$, with
$b_{\bf q}^{\rm A}=u_{\bf q}\alpha_{\bf q}+v_{\bf q}\beta_{-\bf q}^{\dagger}$ and
${b_{\bf q}^{\rm B}}^\dagger=v_{\bf q}\alpha_{\bf q}+u_{\bf q}\beta_{-\bf q}^{\dagger}$
The usual Bogoliubov parameters are given by
\begin{equation}
u_{\bf q} = \Biggl[\frac{1+\kappa_{\bf q}}
            {2\kappa_{\bf q}}\Biggr]^{\frac{1}{2}}, ~~
v_{\bf q} = -\text{sgn}(\gamma_{\bf q})\Biggl[\frac{1-\kappa_{\bf q}}
            {2\kappa_{\bf q}}\Biggr]^{\frac{1}{2}},
\end{equation}
with $\gamma_{\bf q}=(\cos q_x+\cos q_y)/2$
and $\kappa_{\bf q} = (1-\gamma_{\bf q}^2)^{1/2}$.
Substituting the new operators $\alpha_{\bf q}$, $\beta_{\bf q}$ into the Hamiltonian
we arrive at the so-called spin-polaron model:
\begin{eqnarray}
H_t^{\phantom{\dagger}}&=&\frac{zt}{\sqrt{N}}\sum_{{\bf q},{\bf k}}
          \bigl[V({\bf q},{\bf k})
          {f_{\bf k}^{\rm A}}^\dagger f_{{\bf k}-{\bf q}}^{\rm B}
          \alpha_{\bf q}^{\phantom{\dagger}}+ \nonumber \\
     &&  \hspace*{11mm}
          +V(-{\bf q},{\bf k}-{\bf q})
          {f_{\bf k}^{\rm A}}^\dagger f_{{\bf k}-{\bf q}}^{\rm B} \beta_{-{\bf q}}^\dagger
          +\text{h.c.}\bigr] \nonumber\\[2mm]
     &+& t' \sum_{\bf k} ({f_{\bf k}^{\rm A}}^\dagger f_{\bf k}^{\rm A}
             +{f_{\bf k}^{\rm B}}^\dagger f_{\bf k}^{\rm B}) \,,
           \nonumber \\
H_J^{\phantom{\dagger}}&=&\sum_{\bf q}\omega_{\bf q}^0\bigl(
          \alpha_{\bf q}^\dagger\alpha_{\bf q}^{\phantom{\dagger}}+
          \beta_{\bf q}^\dagger\beta_{\bf q}^{\phantom{\dagger}}\bigr).
\end{eqnarray}
Here, $N$ is the number of sites in each sublattice,
and the coordination number is $z=4$.
All sums run over the magnetic Brillouin zone.
The bare spin-wave energy is $\omega_{\bf q}^0=
\frac{zJ}{2} \kappa_{\bf q}$.
The interaction vertex between holes and spin waves is given by
$V({\bf q},{\bf k})=u_{\bf q}\gamma_{{\bf k}-{\bf q}}+v_{\bf q}\gamma_{\bf k}$
and vanishes for ${\bf q} =0$ or ${\bf q} =(\pi,\pi)$.
As a consequence, not only the coupling
between holes and long-wavelength spin fluctuations is important but
also the coupling to short-wavelength spin fluctuations.


\section{Green's functions and Born approximation}
\label{sec:green}

The hole Green's function in the absence of impurities is defined by
$
\mathcal{G}_{\mu\nu}^{\phantom{\dagger}}(\tau,{\bf k})=
                 -\langle T_\tau^{\phantom{\dagger}}
                 f_{\bf k}^\mu(\tau) {f_{\bf k}^\nu}^\dagger(0)\rangle
$,
$\mu,\nu\!=\!{\rm A,B}$, with the Fourier transform
$\mathcal{G}_{\mu\nu}^{\phantom{\dagger}}(i\nu_n,{\bf k})=
                  \int_{0}^{\beta}\text{d}\tau\,
                  \mathcal{G}_{\mu\nu}^{\phantom{\dagger}}(\tau,{\bf k})
                  \,\text{e}^{i\nu_n\tau}$,
where $\nu_n=(2n+1)\pi T$ represents fermionic Matsubara frequencies.
%
The spin-wave Green's function acquires the matrix form
\begin{equation*}
\underline{\mathcal{D}}(\tau,{\bf q})=
               \left[\begin{array}{ll}
                 -\langle T_\tau^{\phantom{\dagger}}
                  \alpha_{\bf q}^{\phantom{\dagger}}(\tau)
                  \alpha_{\bf q}^\dagger(0)\rangle
               & -\langle T_\tau^{\phantom{\dagger}}
                  \alpha_{\bf q}^{\phantom{\dagger}}(\tau)
                  \beta_{\bf -q}^{\phantom{\dagger}}(0)\rangle        \\
                 -\langle T_\tau^{\phantom{\dagger}}
                  \beta_{\bf -q}^\dagger(\tau)
                  \alpha_{\bf q}^\dagger(0)\rangle
               & -\langle T_\tau^{\phantom{\dagger}}
                  \beta_{\bf -q}^\dagger(\tau)
                  \beta_{\bf -q}^{\phantom{\dagger}}(0)\rangle
               \end{array}\right].
\end{equation*}
The unperturbed spin-wave propagator is
\begin{equation}\label{Dsw0}
\underline{\mathcal{D}}_{\,0}^{-1}(i\omega_n,{\bf q}) =
               \left[\begin{array}{cc}
                 i\omega_n-\omega_{\bf q}^0   &  0  \\
                      0  &  - i\omega_n-\omega_{\bf q}^0
               \end{array}\right].
\end{equation}
where $\omega_n=2\pi nT$,
Introducing a self-energy matrix $\underline{\Sigma}(i\omega_n,{\bf q})$,
the solution of the Dyson equation reads
\begin{widetext}
\begin{equation} \label{dyson}
\underline{\mathcal{D}}(i\omega_n,{\bf q})=
               \frac{1}{d(i\omega_n,{\bf q})}
               \left[\begin{array}{cc}
               i\omega_n+\omega_{\bf q}^0 + \Sigma_{22}(i\omega_n,{\bf q})  &
               -\Sigma_{12}(i\omega_n,{\bf q})                                  \\
               -\Sigma_{21}(i\omega_n,{\bf q})                              &
               -i\omega_n+\omega_{\bf q}^0 + \Sigma_{11}(i\omega_n,{\bf q})
               \end{array}\right],
\end{equation}
where
\begin{equation}
d(i\omega_n,{\bf q})=\bigl[i\omega_n+\omega_{\bf q}^0+\Sigma_{22}(i\omega_n,{\bf q})\bigr]
                     \bigl[i\omega_n-\omega_{\bf q}^0-\Sigma_{11}(i\omega_n,{\bf q})\bigr]+
                     \Sigma_{12}(i\omega_n,{\bf q})\Sigma_{21}(i\omega_n,{\bf q}).
\end{equation}
The definition of the spin-wave propagators dictates that
$\Sigma_{22}(i\omega_n,{\bf q})=\Sigma_{11}(-i\omega_n,-{\bf q})$ and
$\Sigma_{12}(i\omega_n,{\bf q})=\Sigma_{21}(i\omega_n,{\bf q})$.

\subsection{Spin-wave self-energies}

The spin-wave self-energies,
$\Sigma_{11}(i\omega_n,{\bf q})$ and
$\Sigma_{12}(i\omega_n,{\bf q})$,
are calculated in the spirit of SCBA,\cite{fulde}
i.e., using the leading (bubble) diagram describing
the decay of the spin fluctuations into
a particle-hole pair with the {\em full} hole propagators,
see e.g. Fig.~1 of Ref.~\onlinecite{horsch}.
This approximation amounts to a summation of an infinite class of non-crossing diagrams;
vertex corrections can be shown to be small for small doping \cite{sherman} and will be
neglected.
Explicitly, the self-energies are given by
\begin{eqnarray}
\label{sigma11}
\Sigma_{11}(i\omega_n,{\bf q}) &=&
                 (zt)^2\frac{1}{\beta}\sum_{i\nu_n}
                 \frac{1}{N}\sum_{{\bf k}}
                 \bigl[V({\bf q},{\bf k})\bigr]^2\,
                 \mathcal{G}_{\rm AA}(i\omega_n+i\nu_n,{\bf k})\,
                 \mathcal{G}_{\rm BB}(i\nu_n,{\bf k}-{\bf q}), \\
\label{sigma12}
\Sigma_{12}(i\omega_n,{\bf q}) &=&
                 (zt)^2\frac{1}{\beta}\sum_{i\nu_n}
                 \frac{1}{N}\sum_{{\bf k}}
                 V({\bf q},{\bf k})\,V(-{\bf q},{\bf k}-{\bf q})\,
                 \mathcal{G}_{\rm AA}(i\omega_n+i\nu_n,{\bf k})\,
                 \mathcal{G}_{\rm BB}(i\nu_n,{\bf k}-{\bf q}).
\end{eqnarray}
For simplicity and as in Ref.~\onlinecite{belkaMag},
in our numerics we will not perform a fully
self-consistent calculation for the hole propagators, but instead use
a trial form for the hole spectral density (derived from the SCBA solution of
the single-hole problem) together with a rigid-band approximation.
(A fully self-consistent SCBA calculation at finite hole concentration has
been performed in Ref.~\onlinecite{sherman} -- it shows that at small hole
concentrations the hole spectrum changes little with doping, i.e.,
the rigid-band picture is reliable.)

\vspace*{2mm}
\end{widetext}

\subsection{Hole spectrum}

The single-hole problem in the $t$-$J$ model has been studied
extensively using various techniques, including the SCBA
within the spin-polaron model sketched above.\cite{rink,kane,martinez}
It has been established that the spectral function for a single hole
consists of a well-defined peak a low energies,
$\rho_{\text{coh}}(\omega,{\bf k})$, which corresponds to
the coherent motion of the dressed hole quasiparticle,
and a broad incoherent part at higher energies,
\begin{equation}\label{holespf}
\rho(\omega,{\bf k})=\rho_{\text{coh}}(\omega,{\bf k})
+\rho_{\text{incoh}}(\omega,{\bf k}) .
\end{equation}

The coherent hole motion arises mainly from a combination of hopping
and spin-flip processes and results in a narrow quasiparticle dispersion, $\varepsilon_{\bf k}$,
with a bandwidth of order $2J$ and a dispersion minimum
at momenta ${\bf k}_\text{min}=(\pm \frac{\pi}{2},\pm \frac{\pi}{2})$.
The quasiparticle weight, $Z_0$, is reduced from unity due to the dressing
of the hole with spin fluctuation and scales with $J/t$ in the
physically relevant parameter regime.
Therefore, for the coherent part we use \cite{horsch,belkaMag}
\begin{equation}\label{coh}
\rho_{\text{coh}}(\omega,{\bf k})=Z_0\,\delta(\omega-\varepsilon_{\bf k})
                 \,\Theta(2J-\varepsilon_{\bf k}),
\end{equation}
with the dispersion $\varepsilon_{\bf k}=J\bigl(\cos^2k_x+\cos^2k_y\bigr)$.
We have assumed that near the minima the dispersion is isotropic
and can be approximated with $\varepsilon_{\bf k}\approx \bar{k}^2/(2m_\text{eff})$,
where $\bar{k}$ measures the distance from the ${\bf k}_\text{min}$.
Consequently, the effective mass of the hole scales as $m_\text{eff}^{-1}=2J$.
Here a remark is in order:
In the $t$-$J$ model without next-neighbor hopping, $t'=0$,
the dispersion near ${\bf k}_\text{min}$
is known to be anisotropic, i.e., rather flat along the $(0,\pi)$--$(\pi,0)$
direction.\cite{tj_disp}
However, photoemission results of Sr$_2$CuO$_2$Cl$_2$ indicate an almost isotropic
dispersion;\cite{arpes}
on the level of the $t$-$J$ model a $t'$ term is needed to capture this physics.\cite{tpr_disp}
Later on, it has been shown that a $t$-$t'$-$J$ model reproduces salient photoemission
features of different cuprate families.\cite{arpes2}
Therefore, we employ an isotropic dispersion in the following;
we have also performed a few calculations with an anisotropic dispersion
appropriate for $t'=0$,
leading to qualitatively similar results
(the $\delta_c$ in Fig.~\ref{fig_disp_diffJ} below changes by less than 10\%).

The incoherent part of the spectrum arises from hole hopping inside
the spin-polaron (or spin-bag) quasiparticle, with the characteristic energy scale $t$.
A reasonable approximation is a momentum-independent constant:
\begin{equation}\label{incoh}
\rho_{\text{incoh}}(\omega)=\frac{1-Z_0}{W}\,\Theta(\omega-2J)\,
                   \Theta(W+2J-\omega),
\end{equation}
where $W= 2zt$ is the incoherent bandwidth.

Within a rigid-band approximation, the hole chemical potential $\mu(\delta)$
is fixed by the hole density $\delta$ through
\begin{equation}\label{chem}
\delta=\frac{1}{N}\sum_{\bf k}\int_{-\infty}^{+\infty} \text{d}\omega\,
               \rho(\omega+\mu,{\bf k})\, n_{\text{F}}(\omega),
\end{equation}
where $\rho(\omega,{\bf k})$ is the doping-independent hole spectral function
\eqref{holespf}, and $n_{\text{F}}(\omega)$ the Fermi function.
For small hole doping ($\delta\ll Z_0$) the chemical potential lies
in the coherent band and at zero temperature has a value $\mu=\pi\delta J/Z_0$.

Eqs.~(\ref{dyson},\ref{sigma11},\ref{sigma12},\ref{chem}) completely describe the
interaction between hole quasiparticles and spin waves in the impurity-free
host and have been used in Ref.~\onlinecite{belkaMag} to investigate the
destruction of magnetism in weakly doped cuprates.


\section{The effect of impurities}
\label{sec:impurity}

In this section we extend the theory to account for a finite density
of non-magnetic impurities in the system.
As discussed in Sec.~\ref{sec:idea}, we assume that the vacancies
primarily act as point-like potential scatterers for the mobile holes.
Neglecting interference effects between spin-wave and impurity scattering processes for holes,
we can capture the scattering off the impurities using a
self-consistent $T$-matrix approach for the {\em dressed} hole quasiparticles,
Eq.~\eqref{holespf}.

\subsection{Scattering potential}

Before proceeding a brief discussion of the quantum chemistry aspects is in order.
Remarkably, there is no consensus on the description of Zn impurities within a
one-band model for the CuO planes.
Zn has a closed $3d^{10}$ configuration, suggesting that holes are expelled.\cite{impreview}
This seems to be in agreement with cluster calculations in Ref.~\onlinecite{quchem1},
which find the environment of a Zn atom to be similar to an undoped parent material.
However, recent LDA calculations \cite{quchem2} have predicted a potential on the Zn site
being repulsive for electrons (i.e.~attractive for holes).
In any case, the impurity potential is likely strong compared to the electron hopping.
In what follows we will model Zn as repulsive point-like scatterer for holes;
for some parameter values we have also performed calculations with an attractive
potential; the results will be mentioned below.
(We note that Ref.~\onlinecite{sushkovLi} assumed a strongly attractive potential
for holes associated to Li impurities in \lclo\ -- if this is correct, and the
potential of Zn is repulsive for holes, this would provide a reasonable explanation
for the different properties of \lclo\ and \lsczo.)

In our calculations, we will neglect the effect of the impurities
on the spin waves.
This piece of physics is well-studied:\cite{sandvik,cherny}
the magnetism is suppressed rather slowly with vacancy doping (compared to hole doping).
In principle, spin-wave scattering off the impurities could be taken into
account along the lines of Ref.~\onlinecite{cherny}.
In Sec.~\ref{sec:mainres} we will account for this dilution/percolation effect by simply
multiplying our results with a magnetization reduction factor taken from the
studies of the diluted quantum Heisenberg antiferromagnet.

Further, we will also neglect the physics of impurity-induced
moments.\cite{indmom1,indmom1b,indmom1c,indmom2,indmom3}
We believe that magnetism arising from these moments alone is too weak
to explain the large observed N\'eel temperatures, i.e., true bulk
magnetism is required.
In a bulk-ordered phase, the effective impurity moments will order below \TN\
due to the mean field arising from the bulk order parameter -- this
has been nicely seen \cite{huecker02} in \lsczo.
However, the contribution of the impurity moments to the total order
parameter is small.

\subsection{Self-consistent $T$-matrix approach}

For small impurity concentration, $n_{\text{imp}}\ll 1$, interference effects between
different impurities are negligible, and the self-consistent $T$-matrix
approximation \cite{hirschfeld} (SCTMA) is suitable to account for multiple scattering
on a single impurity.
Here we employ the SCTMA for spin-polaron quasiparticles, i.e., for holes
which are already dressed by the interaction with spin waves.
Within SCTMA the full hole Green's function
$\tilde{\mathcal{G}}(i\nu_n,{\bf k})$ is given by
\begin{equation}\label{pertGreen}
\tilde{\mathcal{G}}^{-1}(i\nu_n,{\bf k})=
                        \mathcal{G}^{-1}(i\nu_n,{\bf k})
                       -\tilde{\Sigma}(i\nu_n,{\bf k}),
\end{equation}
where $\mathcal{G}(i\nu_n,{\bf k})$ is the hole Green's function
corresponding to the spectral density \eqref{holespf},
and the hole self-energy from impurity scattering is
\begin{equation}
\tilde{\Sigma}(i\nu_n,{\bf k})=n_{\text{imp}}\,T_{{\bf k},{\bf k}}(i\nu_n).
\end{equation}
For point-like scatterers of strength $V_0$ the $T$-matrix depends only on frequency:
\begin{equation}
T_{{\bf k},{\bf k'}}(i\nu_n)\equiv
T(i\nu_n)=\frac{V_0}{1-V_0\,\tilde{\mathcal{G}}(i\nu_n)},
\end{equation}
and $\tilde{\mathcal{G}}(i\nu_n)=
N^{-1}\sum_{\bf k}\tilde{\mathcal{G}}(i\nu_n,{\bf k})$ is the local hole Green's function.

With the replacements $\mathcal{G} \rightarrow \tilde{\mathcal{G}}$ \eqref{pertGreen}
in Eqs.~\eqref{sigma11}, \eqref{sigma12} and
$\rho \rightarrow \tilde{\rho} = -\Im\mathfrak{m}\,\tilde{\mathcal{G}}/\pi$ in
\eqref{chem}
we can now calculate how the impurities affect the
spin-wave spectrum and consequently the magnetic properties of the AF host.
In contrast to Ref.~\onlinecite{horsch} we will not employ
any further approximations.



\section{Magnetic properties}
\label{sec:magprop}

The spin-wave Green's functions defined in Sec.~\ref{sec:green}
allow for a direct calculation of static magnetic properties.

\subsection{Staggered magnetization}

The staggered magnetization in the N\'eel state
decreases with hole doping,
\begin{equation}\label{stagg}
m(\delta)=m_0-\Delta m(\delta),
\end{equation}
where the first term denotes the staggered moment in
2D AF reduced by ground-state spin-wave fluctuations only,
\begin{equation}
m_0=\frac{1}{2}-\frac{1}{2N}\sum_{\bf q} \left(\frac{1}{\kappa_{\bf q}}-1\right),
\end{equation}
with $\kappa_{\bf q} = (1-\gamma_{\bf q}^2)^{1/2}$.
For a square lattice and spin-1/2 we have the well-known
spin-wave result $m_0 \approx 0.303$
(which is in excellent agreement with $m_0 = 0.3070(3)$ from
quantum Monte Carlo simulations \cite{sandvik_m0}).
The effect of doped holes on the AF order is included through the second
term in \eqref{stagg}, given by: \cite{diffeq}
\begin{equation}\label{deltam}
\Delta m(\delta)=\frac{1}{N}\sum_{\bf q}
                 \frac{\langle \alpha_{\bf q}^\dagger
                               \alpha_{\bf q}^{\phantom{\dagger}}\rangle-
                       \gamma_{\bf q}\langle
                               \alpha_{\bf q}^{\phantom{\dagger}}
                               \beta_{-\bf q}^{\phantom{\dagger}}\rangle}
                 {\kappa_{\bf q}}.
\end{equation}
The bosonic occupation numbers in \eqref{deltam}, being zero in the undoped case at $T=0$,
can be calculated from the spin-wave Green's functions as
$\langle \alpha_{\bf q}^\dagger\alpha_{\bf q}^{\phantom{\dagger}}\rangle
= - \mathcal{D}_{11}(\tau\!=\!0^-,{\bf q})$,
$\langle \alpha_{\bf q}\beta_{-\bf q}\rangle
= - \mathcal{D}_{21}(\tau\!=\!0^-,{\bf q})$.
The phase boundary of the antiferromagnetic phase is given by $m(\delta_{\text c})=0$.

Notice that $m(\delta)$ \eqref{stagg} refers to the magnetization
per spin (within the spin-wave approximation).
To obtain the magnetization per site, as measured by bulk probes,
$m(\delta)$ has to be multiplied with the factor $(1-\delta-n_{\text{imp}})$;
this does not change the location of the phase boundary, $\delta_{\text{c}}$.
Of course, this mean-field factor only crudely accounts for the
dilution of the magnetic lattice; in particular, percolation physics cannot
be captured. We will comment on this in Sec.~\ref{sec:perc} below.

\subsection{N\'eel temperature}

The above calculations, performed at finite temperature, allow
to determine the N\'eel temperature, \TN, from
$m(\delta,T=T_\text{N})=0$.
This requires a weak AF coupling, $J_\perp$, along the direction perpendicular to
the CuO plane.
In high-\Tc\ cuprates this inter-layer AF coupling is typically
$J_\perp\sim(10^{-5}-10^{-4})J$.
As a consequence, the spin-wave dispersion has a quasi-2D form which
is given by
\begin{equation}
\widetilde{\omega}_{\bf q}^0=\frac{zJ}{2}
\sqrt{\widetilde{\gamma}_0^2-\widetilde{\gamma}_{\bf q}^2},
\end{equation}
where
\begin{equation}
\widetilde{\gamma}_{\bf q}=\gamma_{\bf q}+\epsilon \gamma_{\bf q}^\perp.
\end{equation}
Here $\gamma_{\bf q}^\perp=\cos q_z$,
and the ratio between inter- and intra-layer coupling is
denoted by $\epsilon=J_\perp/(2J)$.

A drawback of the spin-wave approach is that no sensible critical
behavior is obtained, i.e., the magnetization vanishes linearly as
$T\to T_\text{N}$ (the order parameter critical exponent is $\beta=1$).


\section{Numerical results}
\label{sec:results}

In this section we will present our numerical results for the doping dependence
of the AF order parameter with and without impurities.
Note that all energies are given in units of $t$.
Most calculations are done for a two-dimensional system at zero temperature;\cite{broad}
the finite-temperature results are obtained using a finite interlayer coupling.


\subsection{Hole-doped antiferromagnet without impurities}

First we will discuss the case of the doped AF host without impurities
and compare our results with published ones. \cite{horsch,belkaMag}
Extending these calculations, we will consider different forms
of the hole spectrum to study the influence of both hole dispersion
and hole coherence on the host magnetism.

\subsubsection{``Mobile'' holes}

\begin{figure}[t!]
\centering
\psfrag{m}[][]{$\displaystyle{\frac{m(\delta)}{m(0)}}$}
\psfrag{d}[][]{$\delta$}
\psfrag{j1}[][]{\small $J=0.20$}
\psfrag{j2}[][]{\small $J=0.25$}
\psfrag{j3}[][]{\small $J=0.30$}
\includegraphics{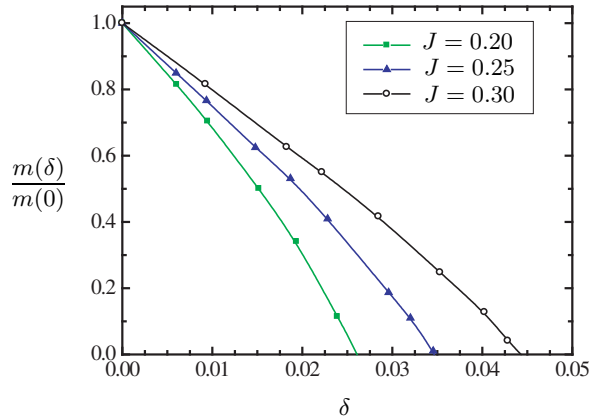}
\caption{
(color online)
Doping dependence of the staggered magnetization
for ``mobile'' (i.e.~dispersive) holes (in the absence of impurities)
for different values of $J$ at $T=0$.
(Typical parameters for cuprates obey $t/J=3\ldots 5$.)
The weight of the dispersive quasiparticle peak in the hole spectrum is $Z_0=0.5\, J/t$;
the lines are guides to the eye only.
The data are very similar to the ones of Ref.~\protect\onlinecite{belkaMag}.
}
\label{fig_disp_diffJ}
\end{figure}

The situation of ``mobile'' holes refers to a hole spectrum function as
obtained from solving the single-hole problem in the $t$-$J$ model, i.e.,
the hole spectral function consists of both a
coherent, dispersive \eqref{coh} and an incoherent, localized \eqref{incoh} part.
In Fig.~\ref{fig_disp_diffJ} the zero-temperature doping dependence of the staggered
magnetization is shown for different values of the coupling $J$ and fixed quasiparticle
weight $Z_0$ (taken from a single-impurity calculation \cite{rink,kane,martinez}).

For $J=0.20$ we obtain the similar critical hole concentration as
Belkasri and Richard, \cite{belkaMag} $\delta_\text{c}\approx 2.7\,\%$.
Our result for $J=0.25$ is a bit smaller than one derived by
Khaliullin and Horsch, \cite{horsch} $\delta_\text{c}\approx4\,\%$,
due to the fact that (as a further approximation) they have taken into account
only the spin-wave self-energies with small momenta.
The disappearance of the AF order at hole concentrations $\delta_\text{c}$
of a few percent is in good agreement with experiments on both \lsco\ and \ybco.

\subsubsection{``Localized'' (incoherent) holes}

Now we turn to a limiting case where the holes are localized in
a certain extended region and only move incoherently
(e.g. around a pinning center).
With the above form of the spectral function \eqref{holespf}
this corresponds to the quasiparticle weight being zero, $Z_0=0$.
Thus, we use $\rho(\omega,{\bf k})=\rho_{\text{incoh}}(\omega)$ \eqref{incoh}.
We study this toy situation to illustrate the
strong influence of hole localization on the magnetic properties.

\begin{figure}[t!]
\centering
\psfrag{m}[][]{$\displaystyle{\frac{m(\delta)}{m(0)}}$}
\psfrag{d}[][]{$\delta$}
\psfrag{j1}[][]{\small $J=0.20$}
\psfrag{j2}[][]{\small $J=0.25$}
\psfrag{j3}[][]{\small $J=0.30$}
\includegraphics{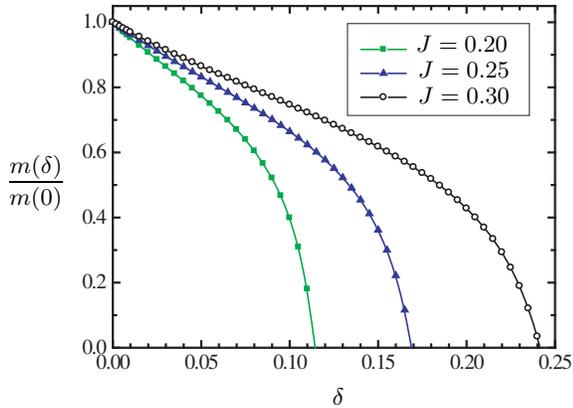}
\caption{
(color online)
Doping dependence of the staggered magnetization now
for ``localized'' holes, i.e., in the absence of a coherent
quasiparticle peak in the hole spectrum ($Z_0=0$).
The calculations have been done at temperature $T=0.005$,
for fixed $J_\perp=0.01\, J$.
}
\label{fig_flat_diffJ}
\end{figure}

One expects the incoherent motion of holes is less destructive for the AF host
in comparison with the case of ``mobile'' holes.
In Fig.~\ref{fig_flat_diffJ} we show the doping dependence of the staggered magnetization
for ``localized'' holes and different values of AF coupling $J$.
($t/J$ is still a relevant parameter here because the bandwidth
of the incoherent motion is controlled by $t$.
Also note that results are shown for finite temperatures and finite coupling $J_\perp$
implying, e.g., that the value of the staggered magnetization $m(0)$ differs from the
$T=0$ value $m_0$. However, this difference is rather small since the effects of $T>0$
and $J_\perp>0$ tend to compensate.)
The critical hole concentrations $\delta_\text{c}$ here are significantly larger
than for dispersive holes, but still much smaller than for strictly static
vacancies (where $\delta_\text{c}$ equals the percolation threshold $z_\text{p} = 40.5\,\%$).
The reason is of course that even our spatially localized holes scramble the
magnetic background in their vicinity.
We call the reader's attention to the analysis of hole mobilities in
Ref.~\onlinecite{hueckerZn}, which concluded that ``completely localized''
holes (i.e.~with zero mobility) still destroy the host magnetism three
times faster than static vacancies -- this gives $\delta_c \approx 13\,\%$
in reasonable agreement with our data in Fig.~\ref{fig_flat_diffJ}.

\begin{figure}[b!]
\centering
\psfrag{m}[][]{$\displaystyle{T_\text{N}(\delta) [t]}$}
\psfrag{d}[][]{$\delta$}
\psfrag{j1}[][]{\small $J=0.20$}
\psfrag{j2}[][]{\small $J=0.25$}
\psfrag{j3}[][]{\small $J=0.30$}
\includegraphics{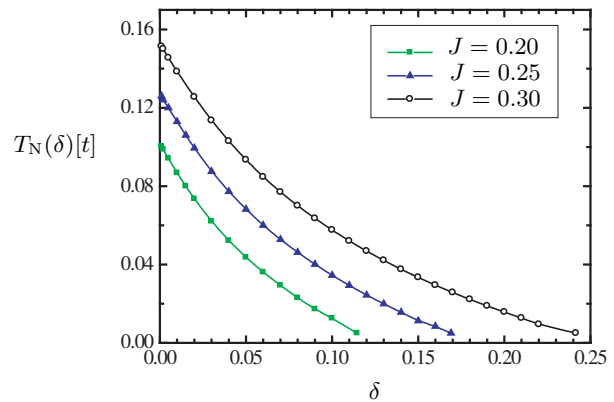}
\caption{
(color online)
Doping dependence of the N\'eel temperature
for ``localized'' holes ($Z_0=0$),
with fixed $J_\perp=0.01\, J$.
(Typical \TN\ values for undoped cuprates are a few hundred kelvins,
e.g., \lco\ has $T_\text{N}\approx 300~$K.
The hopping integral $t$ in the $t$-$J$ model for cuprates
is $300 \ldots 400$ meV $\approx 3000 \ldots 4000~$K.)
}
\label{fig_flat_TNeel}
\end{figure}

In Fig.~\ref{fig_flat_TNeel} the doping dependence of the N\'eel temperature, \TN,
for the ``localized'' holes is depicted for the same set of parameters
as in Fig.~\ref{fig_flat_diffJ}.
As also known for the case of doping with mobile holes,\cite{belkaMag,richardTNeel}
the initial suppression of the N\'eel temperature with doping
is larger than the one of the staggered magnetization.

Comparing the results for ``mobile'' and ``localized'' holes,
Figs. \ref{fig_disp_diffJ} and \ref{fig_flat_diffJ},
one can easily conclude that incoherent motion of holes is less destructive
than coherent one.
Therefore, any mechanism which modifies the hole spectral function in such a way that
suppresses coherent and enhances incoherent motion will lead to a reinforcement of
commensurate antiferromagnetism.
(Note that the limit $t\to 0$, corresponding to immobile holes and percolative
destruction of magnetism, is not described within the current approximation,
but can be captured as in Sec.~\ref{sec:perc} below.)


\subsection{Hole-doped antiferromagnet with non-magnetic impurities}
\label{sec:mainres}

Let us finally turn to our main novel results, namely the magnetic properties
of the system with simultaneous hole and impurity doping.
Here, the spin-wave properties are calculated using a hole spectrum
function corresponding to ``mobile'' holes, i.e., consisting of
a quasiparticle peak and an incoherent background, but
subjected to impurity scattering (which re-distributes weight
in the hole spectrum).

\begin{figure}[t!]
\centering
\centering
\psfrag{m}[][]{$\displaystyle{\frac{m(\delta)}{m(0)}}$}
\psfrag{d}[][]{$\delta$}
\psfrag{j1}[][]{\small $V_0=0\,\,\,\,$}
\psfrag{j2}[][]{\small $V_0=1.0$}
\psfrag{j3}[][]{\small $V_0=2.0$}
\psfrag{j4}[][]{\small $V_0=10.0$}
\hspace*{1mm}
\includegraphics{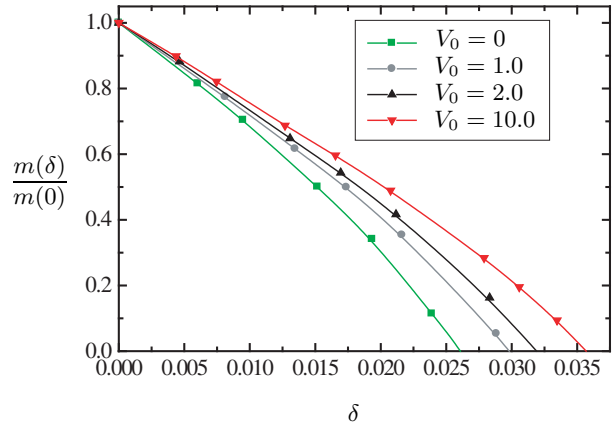}
\caption{
(color online)
Doping dependence of the staggered magnetization
for simultaneous hole and impurity doping.
The different curves correspond to different impurity strengths $V_0$,
the impurity concentration is $n_{\text{imp}}=15\,\%$,
the $V_0=0$ curve is the impurity-free case.
The other parameters are $J=0.20$ and $Z_0=0.5\, J/t$, $T=0$.
}
\label{fig_Znimp_J02}
\end{figure}

Typical results for the AF order parameter as function of the hole doping
are shown in Fig.~\ref{fig_Znimp_J02}.
Clearly, Zn doping increases the critical hole concentration $\delta_\text{c}$.
The difference between the $\delta_\text{c}$ in a case with and
without impurities is of order of one percent which is in reasonable
agreement with experiments \cite{hueckerZn} --
this indicates that we indeed capture the main effect of Zn doping in \lsco.
Notice that $\delta_\text{c}$ increases only weakly
as function of the scattering potential $V_0$ for $V_0>4$.
We have performed calculations for different values of $J$,
with qualitatively similar results
(all $\delta_\text{c}$ values increase for larger $J$, see Fig.~\ref{fig_disp_diffJ}).
In addition, we found that small to moderate negative $V_0$ has qualitatively the
same effect as positive $V_0$
(strong negative $V_0$ leads to hole bound states,
where holes get trapped, and consequently a larger
increase of $\delta_\text{c}$ through impurity doping).

\subsubsection{Less coherent holes}

Within our calculations we can clearly identify the reason for
the antiferromagnetism being more robust in the presence of
impurities.
In the self-consistent $T$-matrix approach of Sec.~\ref{sec:impurity}
the main effect of the impurity scattering is to transfer spectral
weight, i.e., to reduce the weight of the coherent dispersive quasiparticle peak.
We can quantify that: for $n_{\text{imp}}=15\,\%$ and $V_0=2$
the quasiparticle weight $Z_0$ is reduced by roughly $10\,\%$.
A toy calculation with this reduced weight (but in the absence of impurities)
shows that this loss of quasiparticle weight accounts for 3/4 of the
change in $m$ between the clean and impurity-doped cases.
(The remaining 1/4 arises from a change in the quasiparticle dispersion and
from re-shuffling spectral weight at higher energies.)

Thus, the loss of coherence of the hole quasiparticle due to impurity scattering
is the main source of the reinforcement of magnetism.
We can interpret this in real space:
The coherent motion of the hole quasiparticle in the antiferromagnetic background
corresponds to the emission of a spin wave in each hopping step, with a strong
suppression of the order parameter. In contrast, during the incoherent motion
within the quasiparticle the hole also absorbs spin waves (due to
self-retracing paths), and the suppression of antiferromagnetism is less
severe.

\subsubsection{Spin-wave scattering and percolation}
\label{sec:perc}

Our approach has so far neglected the dilution effect of the vacancies on
the magnetic, i.e., spin-wave, part of the model.
On the mean-field level, dilution results in a multiplicative factor
of $(1-\delta-n_\text{imp})$ for $m$ to obtain the staggered magnetization per site.
However, impurity scattering of spin waves is still not included,
and our approach falls short of capturing percolation physics.
While the impurity effect on spin waves could be treated in principle,\cite{cherny}
we choose a different route here:
As dilution of a well-ordered antiferromagnet has been studied
extensively,\cite{sandvik} we can simply multiply our magnetization results
with a renormalization factor $P(n_{\text{imp}})$ which represents the
(normalized) magnetization of a 2D square lattice quantum Heisenberg magnet
diluted with $n_{\text{imp}}$ vacancies -- this quantity can be taken from
quantum Monte Carlo simulations, Fig.~18 of Ref.~\onlinecite{sandvik}.
With this we reproduce a vanishing magnetization at the
percolation threshold.\cite{bilnote}

\subsubsection{Re-appearance of magnetism}

\begin{figure}[t!]
\centering
\psfrag{m}[][]{$\displaystyle{m(\delta)}$}
\psfrag{n}[][]{$\delta$}
\psfrag{j1}[][]{\small $n_\text{imp}=0\quad$}
\psfrag{j2}[][]{\small $n_\text{imp}=15\,\%$}
\includegraphics{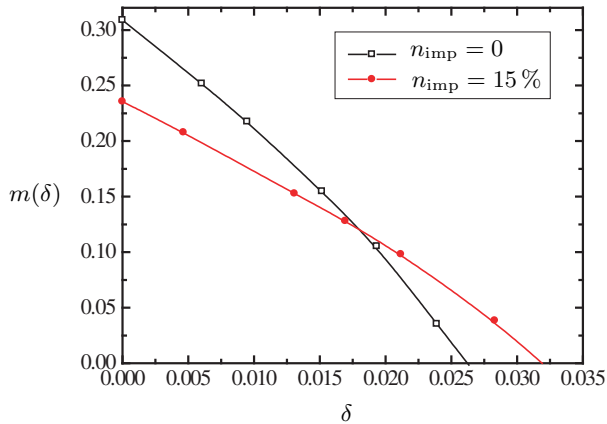}
\caption{
(color online)
Re-appearance of the AF ordering through impurities:
The staggered magnetization,
corrected by the percolation factor $P(n_{\text{imp}})$ (see text),
as function of hole doping for zero and 15\,\% impurity concentration.
The parameters are $J=0.2$, quasiparticle weight $Z_0=0.5\, J/t$,
impurity potential $V_0=2.0$, $T=0$.
}
\label{fig_lscoZn}
\end{figure}

\begin{figure}[t!]
\centering
\psfrag{m}[][]{$\displaystyle{m(\delta)}$}
\psfrag{n}[][]{$n_\text{imp}\,\,[\%]$}
\psfrag{j1}[][]{\small $\delta=2.7\,\%$}
\psfrag{j2}[][]{\small $\delta=2.0\,\%$}
\includegraphics{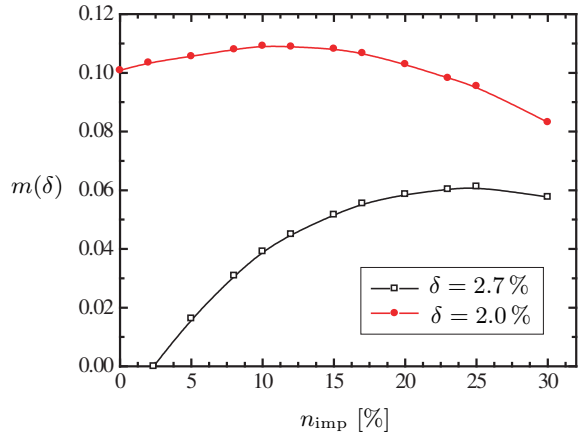}
\caption{
(color online)
Same as Fig.~\protect\ref{fig_lscoZn}, but now showing $m$
as function of the {\em impurity} concentration for
different fixed hole dopings.
$m$ vanishes at the percolation threshold, $n_\text{imp}=40.5\,\%$.
}
\label{fig_nimpZn}
\end{figure}

To emphasize the fact that impurities help the AF order to recover
we show in Fig.~\ref{fig_lscoZn} the staggered magnetization $m$
at zero temperature as a function of hole doping
with and without impurities -- here $m$ is {\em not}
normalized to its value at $\delta=0$.
Another way to show the recovery of long-range order in the presence of
impurities is presented in Fig.~\ref{fig_nimpZn},
where order parameter is plotted as a function
of impurity concentration for two different hole dopings.
For $\delta=2.7\,\%$ the AF order sets in at finite impurity
concentration, $n_\text{imp}\approx 2.5\,\%$.
The presence of impurities causes the re-appearance of the AF order
for hole concentrations greater than $\delta\approx 2.6\,\%$,
and the order parameter depends non-monotonically on the hole doping --
both results are in good agreement with the experiments done on
\lsczo\ by H\"ucker {\it et al.}\cite{hueckerZn}

For a detailed comparison with experiments a few remarks are in order:
(i) The present calculation is strictly valid only for small
impurity concentration, $n_\text{imp}\ll 1$, when the interaction between
impurities is negligible.
(ii) Our simplified treatment of magnetic dilution, utilizing numerical results
for the diluted insulating Heisenberg antiferromagnet, only partially
captures the interplay of dilution and quantum fluctuations.
In particular, in the presence of vacancies and a small amount of holes
magnetic order does not need to persist up to the percolation threshold.\cite{bilnote}
Thus, our data at finite hole doping become unreliable for $n_\text{imp}>25\,\%$.
(iii) Corrections beyond spin-wave theory are required for
the description of critical behavior in the vicinity of $\delta_\text{c}$;
we expect $m$ to vanish with an infinite slope (i.e.~the
order parameter exponent $\beta$ being smaller than unity).
(iv) Experiments usually measure the ordering temperature \TN,
not the zero-temperature order parameter $m$.
Although the two behave in a qualitatively similar fashion,
\TN\ decreases initially faster with doping than $m$.
Also, for a complete picture it may be necessary to include the
interlayer coupling beyond mean-field.\cite{huecker04}
The detailed behavior of \TN\ with both hole and impurity doping,
taking into account percolation-type physics, is beyond the scope of this
paper.

\begin{figure}[b!]
\centering
\psfrag{j1}[][]{\small $V_0=2\,$}
\psfrag{j2}[][]{\small $V_0=10$}
\psfrag{d}[][]{$\delta_{\text{c}}$}
\psfrag{m}[][]{$n_\text{imp}\,\,[\%]$}
\includegraphics{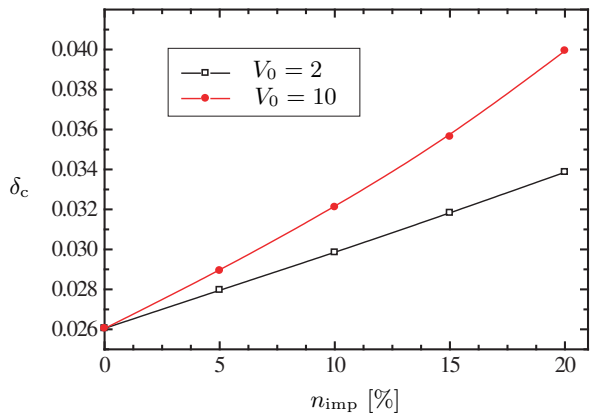}
\caption{
(color online)
Critical hole doping level, $\delta_\text{c}$, as
function of the impurity concentration $n_\text{imp}$
for different values of the scattering potential.
The parameters are $J=0.2$, quasiparticle weight $Z_0=0.5\, J/t$, $T=0$.
}
\label{fig_deltac_nimp}
\end{figure}

In Fig.~\ref{fig_deltac_nimp} we show the evolution of the
critical hole doping level $\delta_\text{c}$ with the impurity concentration.
Interestingly (and in agreement with experiment), the variation
of $\delta_\text{c}$ is rather small, i.e., vacancies cannot
be used to shift $\delta_\text{c}$ to arbitrarily large values.

Last not least we briefly comment on the calculation by Korenblit {\em et al.},\cite{korenblit}
which provides a different explanation for the experiments of Ref.~\onlinecite{hueckerZn}.
They present scaling arguments based on the frustration model, with the key idea
that Zn removes ferromagnetic bonds created by hole doping.
This simple theory predicts that $\delta_\text{c}$ can be increased to
large values by introducing Zn impurities -- this is in contradiction to
the experiment.
Thus, their simple scaling form $\delta \rightarrow \delta(1-n_\text{imp})^2$
cannot apply to larger $\delta$ and $n_\text{imp}$.
Also, the paper makes no reference to the motion and the mobility of holes --
we think that this aspect is crucial for the description of hole-doped cuprates.

\section{Conclusions}
\label{sec:disc}

Motivated by experiments on Zn-doped \lsco,
we have discussed the interplay of mobile holes and static non-magnetic
impurities in lightly doped antiferromagnetic Mott insulators.
Based on a self-consistent spin-wave and $T$-matrix calculations we have shown that
commensurate long-range order is more robust against hole doping
in the presence of non-magnetic impurities as compared to the clean
case.

We believe that the most naive picture, namely that Zn impurities trap holes,
is not appropriate. This can already be seen from the experimental
data: at a hole doping of $2.3\,\%$ the N\'eel temperature depends
strongly and non-monotonically on the impurity concentration
for $4\,\% < n_\text{imp} < 15\,\%$.
Instead, we propose that impurities mainly reduce the hole mobility and
suppress the coherent part of the hole motion (in favor of an incoherent background).
In turn, this suppresses the spin-wave softening caused by coherent hole
motion.
In the case of Ni doping\cite{machiZnNi} this effect is corroborated
by the ordering tendencies of the spin-1 impurity moments.


\acknowledgments

We are grateful to M.~H\"ucker for numerous insightful discussions
about the experiments on Zn-doped \lsco, and for sharing unpublished results.
We also thank
H.~Alloul,
J.~Brinckmann,
C.-H.~Chung,
M.~Indergand,
S.~Florens,
and C.~M.~Varma
for useful comments.
This research was supported by the DFG Center
for Functional Nano\-struc\-tures and the
Virtual Quantum Phase Transitions Institute (Karlsruhe),
as well as NSF Grant No. PHY99-0794 (KITP Santa Barbara).



\begin{thebibliography}{}
\bibitem{diffeq}
Our derivation is closest to the one of Ref.~\protect\onlinecite{belkaMag},
but the equations for the chemical potential and the magnetization correction
differ by factors of 2.
\bibitem{broad}
The zero-temperature calculations have been done with an artificial broadening,
$\eta=0.005$, of $\delta$ peaks in the propagators, to simplify the numerics.
\end{thebibliography}
\end{document}